\documentclass[aps,pra,twocolumn,superscriptaddress,showpacs]{revtex4}
\def\be{\begin{equation}}
\def\ee{\end{equation}}
\def\bea{\begin{eqnarray}}
\def\eea{\end{eqnarray}}
\def\bi{\begin{itemize}}
\def\ei{\end{itemize}}

\usepackage{graphicx}
\usepackage{amsmath}
\usepackage{amssymb}

\begin{document}

\title{ 
        Dynamics of an inhomogeneous quantum phase transition
}

\author{Jacek Dziarmaga}
\author{Marek M. Rams}
\affiliation{ Institute of Physics and 
              Centre for Complex Systems Research, 
              Jagiellonian University,
              Reymonta 4, 30-059 Krak\'ow, 
              Poland}

\begin{abstract}
We argue that in a second order quantum phase transition driven by an inhomogeneous 
quench density of quasiparticle excitations is suppressed when velocity at which a 
critical point propagates across a system falls below a threshold velocity equal to 
the Kibble-Zurek correlation length times the energy gap at freeze-out divided by 
$\hbar$. This general prediction is supported by an analytic solution in the quantum 
Ising chain. Our results suggest, in particular, that adiabatic quantum computers 
can be made more adiabatic when operated in an ``inhomogeneous'' way.
\end{abstract}

\maketitle

%%%%%%%%%%%%%%%%%%%%%%%%%%%%%%%%%%%%%%%%%%%%%%%%%%%%%%%%%%%%%%%%%%%%%%%%%%%%%%%
\section{ INTRODUCTION }
%%%%%%%%%%%%%%%%%%%%%%%%%%%%%%%%%%%%%%%%%%%%%%%%%%%%%%%%%%%%%%%%%%%%%%%%%%%%%%%

A quantum phase transition is a qualitative change in the ground state of 
a quantum system when one of the parameters in its Hamiltonian passes through
a critical point. In a second order transition a continuous change is 
accompanied by a diverging correlation length and vanishing energy gap. The 
vanishing gap implies that no matter how slowly a system is driven 
through the transition its evolution cannot remain adiabatic near the 
critical point. As a result, after the transition the system is excited 
to a state with a finite correlation length $\hat\xi$ whose size shrinks 
with increasing rate of the transition. This scenario, known as Kibble-Zurek (KZ)
mechanism (KZM), was first described in the context of finite temperature 
transitions \cite{K,Z}. Although originally motivated by cosmology \cite{K}, 
KZM at finite temperature was confirmed by numerical simulations of 
the time-dependent Ginzburg-Landau model \cite{KZnum} and successfully tested by 
experiments in liquid crystals \cite{LC}, superfluid helium 3 \cite{He3}, both 
high-$T_c$ \cite{highTc} and low-$T_c$ \cite{lowTc} superconductors, and 
convection cells \cite{ne}. Most recently, spontaneous appearance of vorticity 
during Bose-Einstein condensation driven by evaporative cooling was observed in 
Ref. \cite{Anderson}. However, the quantum zero temperature limit, which is in 
many respects qualitatively different, remained unexplored until recently, see
e.g. Refs. \cite{3sites,Bodzio1,KZIsing,Dziarmaga2005,Polkovnikov,Levitov,JDrandom,
Cucchietti,Bodzioferro,Hindusi,others}. The recent interest is motivated in part by 
adiabatic quantum computation or adiabatic quantum state preparation, where one would 
like to cross a quantum critical point as adiabatically as possible, and in part by 
condensed matter physics of ultracold atoms, where it is easy to manipulate parameters 
of a Hamiltonian in time and which, unlike their solid state physics counterparts, are 
fairly well isolated from their environment. In fact, an instantaneous quench
to the ferromagnetic phase in a spinor BEC resulted in finite-size ferromagnetic
domains whose origin was attributed to KZM \cite{ferro}. However, since the transition
rate was effectively infinite in that experiment, the KZ scaling relation between the 
average domain size $\hat\xi$ and the quench rate has not been verified. 

The KZM argument is briefly as follows \cite{Z,KZIsing}. When a transition is driven 
by varing a parameter $g$ in the Hamiltonian across an isolated critical point at 
$g_c$, then we can define a dimensionless distance from the critical point as
\be
\epsilon~=~\frac{g-g_c}{g_c}~.
\ee
When $\epsilon\to0$ the correlation length $\xi$ in the ground state diverges as 
$\xi\sim|\epsilon|^{-\nu}$, and the energy gap $\Delta$ between the ground state
and the first excited state vanishes as $\Delta\sim|\epsilon|^{z\nu}$. Setting 
$\hbar=1$ from now on, a diverging $\Delta^{-1}\sim|\epsilon|^{-z\nu}$ is the 
shortest time scale on which the ground state can adjust adiabatically to varying 
$\epsilon(t)$. A generic $\epsilon(t)$ can be linearized near the critical point
$\epsilon=0$ as
\be
\epsilon(t)~\approx~-\frac{t}{\tau_Q}~+~{\cal O}(t^2),
\label{tauQ}
\ee
where the coefficient $\tau_Q$ is called a quench time. Assuming that the system
was initially prepared in its ground state, its adiabatic evolution fails at a 
$\hat\epsilon$ when the time $\hat t$ left to crossing the critical point equals 
the shortest time scale $\Delta^{-1}$ on which the ground state can adjust. Solving 
this equality, we obtain
\bea
\hat\epsilon &\sim& \tau_Q^{-\frac{1}{z\nu+1}}~,   \label{hatepsilon}\\ 
\hat t       &\sim& \tau_Q^{\frac{z\nu}{z\nu+1}}~. \label{hatt}      
\eea
From $\hat\epsilon$ the evolution becomes impulse, i.e. the state does not evolve
but remains frozen in the ground state at $\hat\epsilon$, until 
$-\hat\epsilon$ when the evolution becomes adiabatic again. In this way, 
the ground state at $\hat\epsilon$ with a KZ correlation length
\be
\hat\xi ~\sim~ \hat\epsilon^{-\nu} ~\sim~ \tau_Q^{\frac{\nu}{z\nu+1}}
\label{hatxi}
\ee
becomes the initial excited state for the adiabatic evolution after $-\hat\epsilon$. 
In particular, $\hat\xi^{-1}$ determines density of quasiparticles excited during
the phase transition 
\be
d ~\sim~ \tau_Q^{-\frac{D\nu}{z\nu+1}}
\label{dKZgeneral}
\ee
in $D$ dimensions. Note that when $\tau_Q$ is large, then $\hat\epsilon$ is small 
and the linearization in Eq. (\ref{tauQ}) is self-consistent because the KZM physics 
happens very close to the critical point between $-\hat\epsilon$ and $+\hat\epsilon$.

%%%%%%%%%%%%%%%%%%%%%%%%%%%%%%%%%%%%%%%%%%%%%%%%%%%%%%%%%%%%%%%%%%%%%%%%%%%%%%%
\section{ Inhomogeneous transition }\label{inhom_general}
%%%%%%%%%%%%%%%%%%%%%%%%%%%%%%%%%%%%%%%%%%%%%%%%%%%%%%%%%%%%%%%%%%%%%%%%%%%%%%%

As pointed out in the finite temperature context \cite{Volovik}, in a realistic 
experiment it is difficult to make $\epsilon$ exactly homogeneous throughout a 
system. For instance, in the superfluid $^3He$ experiments \cite{He3} the
transition was caused by neutron irradiation of helium 3. Heat released in each
fusion event, $n~+~^3He~\to~^4He$, created a bubble of normal fluid above the 
superfluid critical temperature $T_c$. As a result of quasiparticle diffusion, 
the bubble was expanding and cooling with a local temperature 
$T(t,r)=\exp(-r^2/2Dt)/(2\pi Dt)^{3/2}$,
where $r$ is a distance from the center of the bubble and $D$ is a diffusion 
coefficient. Since this $T(t,r)$ is hottest in the center, the transition 
back to the superfluid phase, driven by an inhomogeneous parameter 
\be
\epsilon(t,r)~=~\frac{T(t,r)-T_c}{T_c}~,
\ee 
proceeded from the outer to the central part of the bubble with a critical front 
$r_c(t)$, where $\epsilon=0$, shrinking with a finite velocity $v=dr_c/dt<0$.

A similar scenario is likely in the ultracold atom gases in magnetic/optical traps. 
The trapping potential results in an inhomogenous density of atoms $\rho(\vec r)$
and, in general, a critical point $g_c$ depends on atomic density $\rho$. Thus 
even a transition driven by a perfectly uniform $g(t)$ will be effectively 
inhomogeneous,
\be
\epsilon(t,\vec r)~=~
\frac{g(t)-g_c[\rho(\vec r)]}{g_c[\rho(\vec r)]}~,      
\ee
with the surface of critical front, where $\epsilon=0$, moving with a finite velocity. 

According to KZM, in a homogeneous symmetry breaking transition, a state after the 
transition is a mosaic of finite ordered domains of average size $\hat\xi$. Within 
each finite domain orientation of the order parameter is constant but uncorrelated to
orientations in other domains. In contrast, in an inhomogeneous symmetry breaking 
transition \cite{Volovik}, the parts of the system that cross the critical point 
earlier may be able to communicate their choice of orientation of the order parameter
to the parts that cross the transition later and bias them to make the same choice. 
Consequently, the final state may be correlated at a range longer than $\hat\xi$, or 
even end up being a ground state. In other words, the final density of excited 
quasiparticles may be lower than the KZ estimate in Eq. (\ref{dKZgeneral}) or even 
zero. 

From the point of view of testing KZM, this inhomogeneous scenario, when relevant, may 
sound like a negative result: an imperfect inhomogeneous transition supresses KZM.
However, from the point of view of adiabatic computation or adiabatic state preparation
it is the KZM itself that is a negative result: no matter how slow the homogeneous
transition is there is a finite density of excitations (\ref{dKZgeneral}) which decays 
only as a small power of transition time $\tau_Q$. From this perspective, the
inhomogeneous transition may be a way to suppress KZ excitations
and prepare the desired final ground state adiabatically.

To estimate when the inhomogenuity may actually be relevant, in a similar way as in 
Eq. (\ref{tauQ}), we linearize the parameter $\epsilon(t,n)$ in both $n$ and $t$
near the critical front where $\epsilon(t,n)=0$:
\be
\epsilon(t,n)~\approx~\alpha~(n-vt)~.
\label{alpha}
\ee
Here $n$ is position in space, e.g. lattice site number, $\alpha$ is a slope of 
the quench and $v$ is velocity of the critical front. When observed locally at 
a fixed $n$, the inhomogeneous quench in Eq. (\ref{alpha}) looks like the homogeneous 
quench in Eq. (\ref{tauQ}) with
\be
\tau_Q~=~\frac{1}{\alpha v}~.
\label{tauQalphav}
\ee    
The part of the system where $n<vt$, or equivalently $\epsilon(t,n)<0$, is already
in the broken symmetry phase. The orientation of the order parameter chosen in this 
part can be communicated across the critical point not faster than a threshold velocity
\be
\hat v ~\simeq~ \frac{\hat\xi}{\hat t}~.
\label{hatv}
\ee 
When $v\gg\hat v$ the communication is too slow for the inhomogenuity to be relevant, 
but when $v\ll\hat v$ we can expect the final state to be less excited than predicted 
by KZM. 

Given the relation (\ref{tauQalphav}), the condition (\ref{hatv}) can be solved either 
as
\bea
\hat v &\sim& \tau_Q^{-\frac{(z-1)\nu}{z\nu+1}} ~, \label{hatvtauQ}\\
\hat v &\sim& \alpha^{\frac{\nu(z-1)}{1+\nu}}   ~, \label{hatvalpha}
\eea
or as a relation between the threshold transition time and the slope,
\be
\hat\tau_Q~\sim~\alpha^{-\frac{z\nu+1}{1+\nu}}~.
\label{hattauQ}
\ee
This relation means that, for a given inhomogenuity $\alpha$, the transition is
effectively homogeneous when $\tau_Q\ll\hat\tau_Q$, but the inhomogenuity becomes
relevant when the transition is slow enough, $\tau_Q\gg\hat\tau_Q$. In the homogeneous 
limit $\alpha\to0$, the threshold transition time $\hat\tau_Q\to\infty$.

The threshold velocity in Eq. (\ref{hatv}) appeared for the first time in the context 
of finite temperature classical phase transitions \cite{Volovik} where it looks formally 
the same, but the underlying physics is qualitatively different: the scales $\hat\xi$ 
and $\hat t$ are determined not by the gap of a quantum Hamiltonian, but the relaxation 
time of an open classical system. Nevertheless, the key mechanism that non-zero 
order parameter penetrates from the symmetry broken phase into the symmetric phase ahead 
of the critical front seems to be the same. 

In the next Section we rederive results (\ref{hatvtauQ},\ref{hatvalpha},\ref{hattauQ})
from a different perspective.

%%%%%%%%%%%%%%%%%%%%%%%%%%%%%%%%%%%%%%%%%%%%%%%%%%%%%%%%%%%%%%%%%%%%%%%%%%%%%%%
\section{ KZM in space }\label{KZMinspace_general}
%%%%%%%%%%%%%%%%%%%%%%%%%%%%%%%%%%%%%%%%%%%%%%%%%%%%%%%%%%%%%%%%%%%%%%%%%%%%%%%

References \cite{Dorner,Damski} considered a ``phase transition is space'' 
where $\epsilon(n)$ is inhomogeneous but time-independent. In the same 
way as in Eq. (\ref{alpha}), this parameter can be linearized in $n-n_c$, 
\be
\epsilon(n)~\approx~\alpha~(n-n_c)~,
\label{nc}
\ee
near the static critical front at $n=n_c$ where $\epsilon=0$. 
The system is in the broken symmetry phase where $n<n_c$ and in the symmetric
phase where $n>n_c$. In the first ``local approximation'', we would expect that 
the order parameter behaves as if the system were locally  uniform: it is nonzero 
for $n<n_c$ only, and tends to zero as $(n_c-n)^\beta$ when $n\to n_c^-$ with
the critical exponent $\beta$. However, this first approximation is in contradiction 
with the basic fact that the correlation (or healing) length $\xi$ diverges as
$\xi\sim|\epsilon|^{-\nu}$ near the critical point and the diverging $\xi$ is 
the shortest length scale on which the order paramater can adjust to (or heal 
with) the changing $\epsilon(n)$. Consequently, when approaching $n_c^-$ the local
approximation $(n_c-n)^\beta$ must break down when a local correlation length
$\xi\sim[\alpha(n_c-n)]^{-\nu}$ equals the distance remaining to the critical point
$(n_c-n)$. Solving this equality with respect to $\xi$, we obtain 
\be
\hat\xi\sim\alpha^{-\nu/(1+\nu)}~.
\label{tildexigeneral}
\ee 
From $n-n_c\simeq-\hat\xi$ the evolution
of the order parameter in $n$ becomes ``impulse'', i.e, the order parameter does 
not change until $n-n_c\simeq+\hat\xi$ in the symmetric phase where it begins
to follow the local $\epsilon(n)$ again and quickly decays to zero on the same length
scale of $\hat\xi$. This ``KZM in space'' predicts that a nonzero order parameter 
penetrates into the symmetric phase to a depth
\be
\delta n ~\sim~ \hat\xi ~\sim~ \alpha^{-\frac{\nu}{1+\nu}}~.
\label{deltangeneral}
\ee 
The critical point is effectively ``rounded off'' on the length scale of $\hat\xi$.
As a consequence, we expect a non-zero gap scaling as
\be
\hat\Delta~\sim~\hat\xi^{-z}~\sim~\alpha^{\frac{z\nu}{1+\nu}}~
\label{tildeDeltageneral}
\ee
as opposed to the local approximation, where we would expect gapless 
quasiparticles near the critical point. 

We expect the finite gap in Eq. (\ref{tildeDeltageneral}) to prevent excitation 
of the system even when the critical point $n_c$ in Eq. (\ref{slope}) moves with 
a finite velocity, $n_c(t)=vt$, up to a threshold velocity
\be
\hat v~\sim~
\frac{\hat\xi}{\hat\Delta^{-1}}~\sim~
\alpha^{\frac{\nu(z-1)}{1+\nu}}
\label{tildevgeneral}
\ee 
which is identical with the $\hat v$ in Eq. (\ref{hatvalpha}). 

In the following Sections we test these predictions in the quantum Ising chain.

%%%%%%%%%%%%%%%%%%%%%%%%%%%%%%%%%%%%%%%%%%%%%%%%%%%%%%%%%%%%%%%%%%%%%%%%%%%%%%%
\section{ Quantum Ising chain } 
%%%%%%%%%%%%%%%%%%%%%%%%%%%%%%%%%%%%%%%%%%%%%%%%%%%%%%%%%%%%%%%%%%%%%%%%%%%%%%%

The model is 
\be
H~=
~-~\sum_{n=1}^N     g_n~\sigma^x_n 
~-~\sum_{n=1}^{N-1} \sigma^z_n\sigma^z_{n+1} ~.
\label{HS}
\ee 
For $N\to\infty$, a uniform system with $g_n=g$ has two critical points at 
$g=\pm1$ separating a ferromagnetic phase, when $|g|<1$, from two paramagnetic 
phases, when $|g|>1$. We focus on the critical point at $g=1$ when $\epsilon=g-1$. 
Given $z=1$ and $\nu=1$, we expect
\be
\hat v~\simeq~1
\label{hatv1}
\ee
independent of either $\tau_Q$ or $\alpha$. The quench is exactly solvable 
for homogeneous $g$ \cite{Dziarmaga2005,Levitov}, but even in an inhomogenous case
some useful analytic insights can be obtained as follows.
  
After Jordan-Wigner transformation to spinless fermionic operators $c_n$, 
$\sigma^x_n=1-2 c^\dagger_n c_n$ and 
$\sigma^z_n=-\left( c_n + c_n^\dagger \right)\prod_{m<n}(1-2 c^\dagger_m c_m)$,
Eq. (\ref{HS}) becomes 
\bea
H ~=~ 2 \sum_{n=1}^N g_n ~c_n^\dagger  c_n-
\sum_{n=1}^{N-1}
\left( 
c_n^\dag c_{n+1} + c_{n+1} c_n + {\rm h.c.}
\right)
\label{Hpm}
\eea
This quadratic $H$ is diagonalized to $H=\sum_m\omega_m\gamma_m^\dag\gamma_m$ by 
a Bogoliubov transformation 
$c_n=\sum_{m=0}^{N-1}(u_{nm}\gamma_m + v^*_{nm} \gamma_m^\dagger)$ with
$m$ numerating $N$ eigenmodes of stationary Bogoliubov-de Gennes equations
\bea
\omega_m u_{n,m}^\pm = 2 g_n u_{n,m}^\mp - 2 u_{n\mp 1,m}^\mp ~
\label{BdG}
\eea
with $\omega_m\geq0$. Here $u_{nm}^\pm\equiv u_{nm} \pm v_{nm}$.

%%%%%%%%%%%%%%%%%%%%%%%%%%%%%%%%%%%%%%%%%%%%%%%%%%%%%%%%%%%%%%%%%%%%%%%%%%%%%%%%%%%
\begin{figure}
\includegraphics[width=0.99\columnwidth,clip=true]{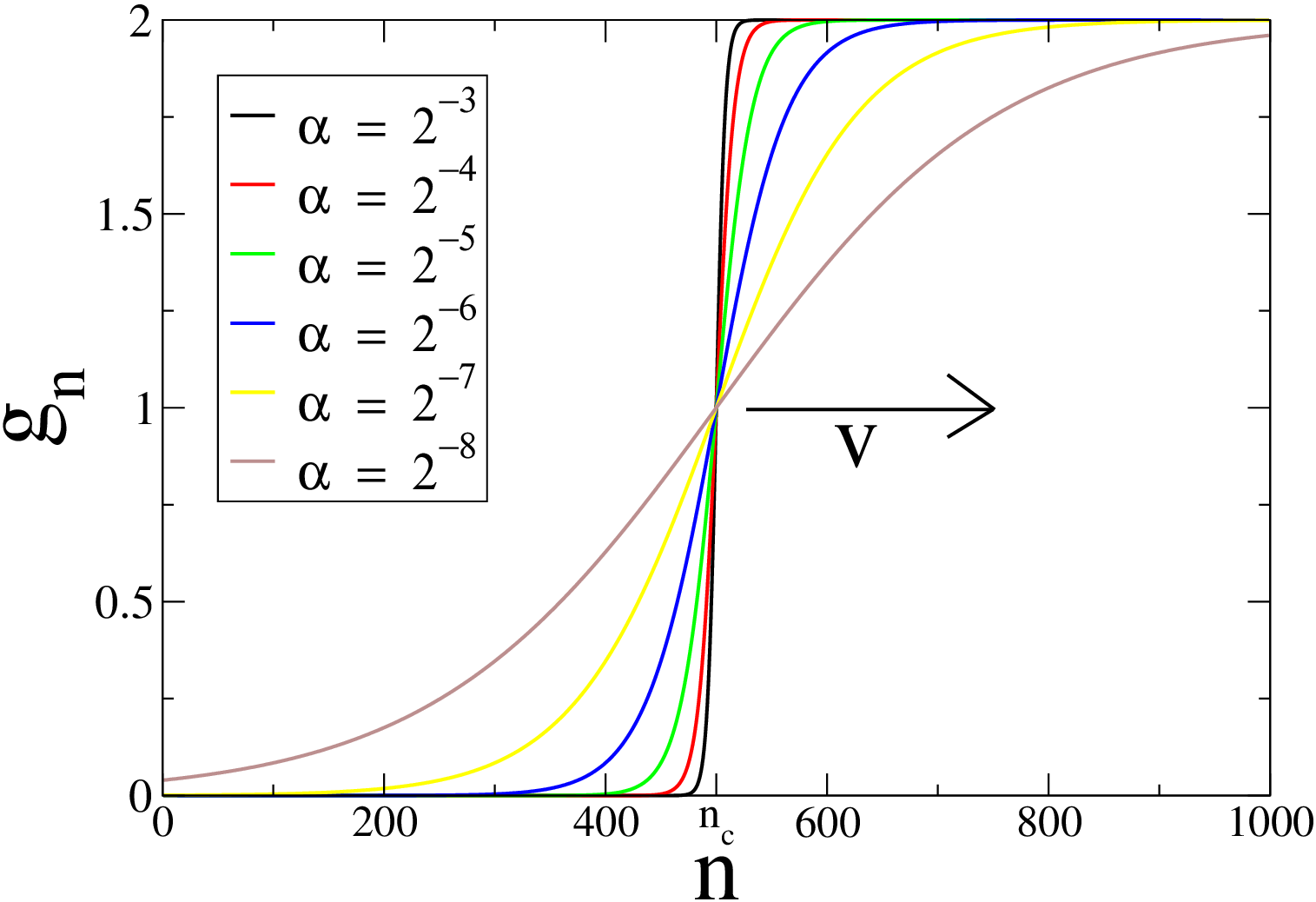}
\caption{ 
The critical front in Eqs. (\ref{slant},\ref{slantv}).
}
\label{Figslant}
\end{figure}
%%%%%%%%%%%%%%%%%%%%%%%%%%%%%%%%%%%%%%%%%%%%%%%%%%%%%%%%%%%%%%%%%%%%%%%%%%%%%%%%%%%%%

%%%%%%%%%%%%%%%%%%%%%%%%%%%%%%%%%%%%%%%%%%%%%%%%%%%%%%%%%%%%%%%%%%%%%%%%%%%%%%%
\section{ Ising chain: KZM in space }
%%%%%%%%%%%%%%%%%%%%%%%%%%%%%%%%%%%%%%%%%%%%%%%%%%%%%%%%%%%%%%%%%%%%%%%%%%%%%%%

To begin with, we consider the ground state of the quantum Ising chain in
a static inhomogeneous tranverse field $g_n$ which can be linearized
near the critical point $g=1$ as 
\be
\epsilon(n)~=~g_n~-~1~\approx~\alpha~(n-n_c)~,
\label{slope}
\ee
compare with Eq. (\ref{nc}). The chain is in the (broken symmetry) ferromagnetic 
phase where $n<n_c$ and in the (symmetric) paramagnetic phase where $n>n_c$. We 
want to know if the nonzero ferromagnetic magnetization $Z_n=\langle\sigma^z_n\rangle$ 
in the ferromagnetic phase penetrates across the critical point into
the paramagnetic phase and what is the depth $\delta n$ of this 
penetration.

Since in a homogeneous system quasiparticle spectrum is gapless at the critical
point only, we expect low energy quasiparticle modes $u^\pm_{n,m}$ to be localized 
near the critical point at $n_c$ where we can use the linearization in Eq. (\ref{slope}). 
We also expect that these low energy modes are smooth enough to treat $n$ as continuous 
and make a long wavelength approximation
\be
u_{n\mp 1,m}^\mp \approx u_{n,m}^\mp \mp \frac{\partial}{\partial n}u_{n,m}^\mp
\label{LWA}
\ee
in Eq. (\ref{BdG}). Under these assumptions, we obtain a long-wavelength equation
\bea
\omega_m u_m^\pm = 2\alpha(n-n_c) u_m^\mp \pm 2 \partial_n u_m^\mp ~.
\label{lowBdG}
\eea
After some algebra, its eigenmodes can be found as  
\bea
u_m(n)   &\propto& \psi_{m-1}(x) + \psi_{m}(x) ~,\nonumber\\
v_m(n)   &\propto& \psi_{m-1}(x) - \psi_{m}(x) ~,\nonumber\\
\omega_m &   =   & \sqrt{8m\alpha}~,
\label{spectrum}
\eea
where 
\be
x~=~\sqrt{\alpha}(n-n_c)~,
\label{x}
\ee
is a rescaled position, $\psi_{m\geq0}(x)$ are eigenmodes of a harmonic 
oscillator satisfying
\be
\frac12
(-\partial_x^2+x^2)
\psi_m(x)=
(m+1/2)
\psi_m(x)~,
\ee
and $\psi_{-1}(x)=0$. As expected, the modes in Eq. (\ref{spectrum}) are 
localized near $n=n_c$ where $x=0$. A typical width of the lowest energy
eigenmodes is $\delta x\simeq1$, or equivalently 
\be
\delta n~\simeq~\alpha^{-1/2}~.
\label{deltan}
\ee
When $\alpha\ll1$ then $\delta n\gg1$ and the long wavelength approximation 
in Eqs. (\ref{LWA},\ref{lowBdG}) is self-consistent. Thus $\delta n$ in 
Eq. (\ref{deltan}) is the relevant scale of length near $n_c$ and we expect
that this $\delta n$ determines the penetration depth of the spontaneous 
ferromagnetic magnetization into the paramagnetic phase. 

We test this prediction by a numerical solution for an inhomogeneous
transverse magnetic field 
\be
g_n~=~1~+~\tanh\left[\alpha(n-n_c)\right]~,
\label{slant}
\ee
which is shown in Fig. (\ref{Figslant}) with a variable slope $\alpha$.
This field can be self-consistenly linearized near $n=n_c$ as in Eq. (\ref{slope})
because, when the slope $\alpha\ll1$, the predicted $\delta n\simeq\alpha^{-1/2}$ 
is much shorter than the width $\alpha^{-1}$ of the $\tanh$. 

\begin{figure}
\includegraphics[width=0.99\columnwidth,clip=true]{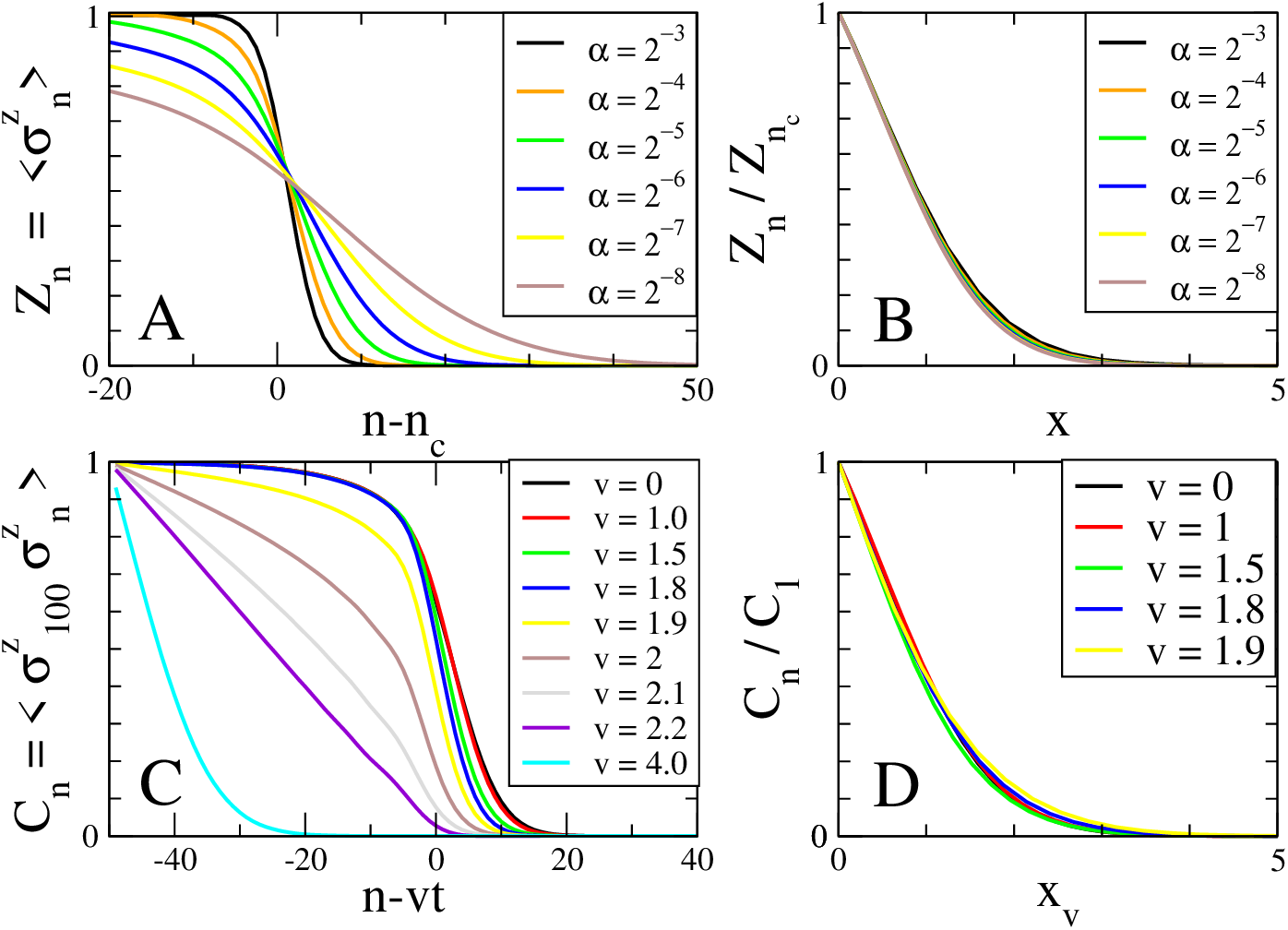}
\caption{ 
In A and B, exact numerical spontaneous magnetization as a function of $n-n_c$ and 
$x=\sqrt{\alpha}(n-n_c)$ respectively. 
In C and D, ferromagnetic correlation between the site $100$ in the ferromagnetic 
phase and a site $n$ when $n_c=vt=150$ as a function of $n-vt$ and $x_v$ in Eq. (\ref{xv}) 
respectively. Results in C and D were obtained with the Vidal algorithm \cite{Vidal} for 
$\alpha=2^{-5}$ and $N=256$. When $v\gg2$ there is no ferromagnetic correlation across the 
critical point at $n-vt=0$, see C, and when $v\ll2$ the correlation penetrates into the 
paramagnetic phase to a depth of $\delta x_v\simeq1$, see D. 
} 
\label{Figpenetration}
\end{figure}

Figures {\ref{Figpenetration}}A and B show how the spontaneous ferromagnetic 
magnetization $Z_n=\langle\sigma_n^z\rangle$ from the ferromagnetic phase, 
where $n<n_c$, penetrates into the paramagnetic phase, where $n>n_c$. 
In particular, the collapse of the rescaled plots in Fig. \ref{Figpenetration}B 
demonstrates that the penetration depth is $\delta x\simeq 1$ equivalent to 
$\delta n\simeq\alpha^{-1/2}$~, as predicted in Eqs. 
(\ref{deltangeneral},\ref{deltan}) and Ref. \cite{Dorner}. Paramagnetic spins 
near the critical point are biased towards the direction of spontaneous 
magnetization chosen in the ferromagnetic phase. 

Moreover, the analytic solution (\ref{spectrum}) implies a finite (relevant) gap
\be
\hat\Delta~=~\omega_0~+~\omega_1~=~\sqrt{8\alpha}~
\ee
in accordance with the scaling $\sim\alpha^{1/2}$ predicted by the general 
Eq. (\ref{tildeDeltageneral}) and the numerics in Ref. \cite{Dorner}. This gap 
is the energy of the lowest relevant (even parity) excitation of two quasiparticles.

%%%%%%%%%%%%%%%%%%%%%%%%%%%%%%%%%%%%%%%%%%%%%%%%%%%%%%%%%%%%%%%%%%%%%%%%%%%%%%%%%%%%%
%%%%%%%%%%%%%%%%%%%%%%%%%%%%%%%%%%%%%%%%%%%%%%%%%%%%%%%%%%%%%%%%%%%%%%%%%%%%%%%%%%%

%%%%%%%%%%%%%%%%%%%%%%%%%%%%%%%%%%%%%%%%%%%%%%%%%%%%%%%%%%%%%%%%%%%%%%%%%%%%%%%
\section{ Ising chain: inhomogeneous transition }
%%%%%%%%%%%%%%%%%%%%%%%%%%%%%%%%%%%%%%%%%%%%%%%%%%%%%%%%%%%%%%%%%%%%%%%%%%%%%%%

Let the critical front in Eq. (\ref{slope}) and Fig. \ref{Figslant} move with
a velocity $v>0$:
\be
n_c(t)~=~vt~.
\ee
A $t$-dependent version of the long-wavelength Eq. (\ref{lowBdG}),
\bea
i\partial_t 
\left(
\begin{array}{c}
u^+ \\
u^-
\end{array}
\right)
~=~
\left[
2\alpha(n-vt) \sigma^x +
2i \sigma^y \partial_n
\right]
\left(
\begin{array}{c}
u^+ \\
u^-
\end{array}
\right)~,
\label{tBdG}
\eea
can be solved exactly for both $v<2$ and $v>2$ with qualitatively different solutions 
in the two regimes. Not incidentally, $v=2$ is the maximal velocity of quasiparticles
at the critical point whose dispersion is $\omega=2|k|$ for small $|k|\ll\pi$.

%%%%%%%%%%%%%%%%%%%%%%%%%%%%%%%%%%%%%%%%%%%%%%%%%%%%%%%%%%%%%%%%%%%%%%%%%%%%%%%
\subsection{ Case of $v<2$ }
%%%%%%%%%%%%%%%%%%%%%%%%%%%%%%%%%%%%%%%%%%%%%%%%%%%%%%%%%%%%%%%%%%%%%%%%%%%%%%%

When $v<2$ equation (\ref{tBdG}) has solutions
\bea
u_m(t,n)
&\propto&
e^{-i\omega_m t}~
\left[
\psi_{m-1}(x_v)
+
e^{i\varphi}\psi_m(x_v)
\right]
e^{ivx_v\sqrt{\frac{m}{2}}}
~,\nonumber\\
v_m(t,n)
&\propto&
e^{-i\omega_m t}~
\left[
e^{i\varphi}\psi_{m-1}(x_v)
-
\psi_m(x_v)
\right]
e^{ivx_v\sqrt{\frac{m}{2}}}
~,\nonumber\\
\omega_m &=&
\left(1-\frac{v^2}{4}\right)^{3/4}~
\sqrt{8\alpha~m}~,
\label{omegam}
\eea
where $m=0,1,2,...$, the phase $\varphi=\arcsin(v/2)/2$, and
\be
x_v~=~
\left(1-\frac{v^2}{4}\right)^{-1/4}~  
\sqrt{\alpha}~ 
(n-vt)~
\label{xv}
\ee
is a rescaled position. When $v\to0$ we recover the static solutions (\ref{spectrum}). 
In the reference frame of $x_v$, which is co-moving with the critical point, the 
solutions (\ref{omegam}) are stationary modes with $\omega_m\geq0$ so there are no 
quasiparticles in the system,
\be
d(v<2)~=~0~,
\ee 
and, in particular, no kinks where $g_n=0$.

As shown in Figs. \ref{Figpenetration}C and D, ferromagnetic correlations penetrate
across the critical point into the paramagnetic phase to a depth $\delta x_v\simeq1$
equivalent to
\be
\delta n_v ~\simeq~ 
\left(1-\frac{v^2}{4}\right)^{1/4}~
\alpha^{-1/2}~. 
\ee
The penetration depth $\delta n_v$ shrinks to $0$ when $v\to2^-$ suggesting 
communication problems across the critical point when $v>2$. 

The same $\delta n_v$ is a typical width of the lowest eigenmodes in the
spectrum (\ref{omegam}). As it shrinks to $0$ when $v\to 2^-$, the eigenmodes
become inconsistent with the long-wavelength approximation in Eq. (\ref{tBdG}).

%%%%%%%%%%%%%%%%%%%%%%%%%%%%%%%%%%%%%%%%%%%%%%%%%%%%%%%%%%%%%%%%%%%%%%%%%%%%%%%
\subsection{ Case of $v>2$ } 
%%%%%%%%%%%%%%%%%%%%%%%%%%%%%%%%%%%%%%%%%%%%%%%%%%%%%%%%%%%%%%%%%%%%%%%%%%%%%%%

When $v>2$ then equation (\ref{tBdG}) can be mapped to a homogeneous transition. 
Indeed, we replace
\be
\tilde t~=~
\left(1-\frac{4}{v^2}\right)^{-1}
\left(t-\frac{n}{v}\right)~,~~
\tilde n~=~n~,
\label{prime}
\ee
introducing local time $\tilde t$ measured from the moment the critical point passes 
through $n$, and simultaneously make a transformation 
\bea
\left(
\begin{array}{c}
u^+ \\
u^-
\end{array}
\right) &=&
\left(
\begin{array}{cc}
\sqrt{1-\frac{4}{v^2}} & \frac{2i}{v}  \\
0                      & 1
\end{array}
\right)
\left(
\begin{array}{c}
\tilde u^+\\
\tilde u^-
\end{array}
\right)~
\eea
bringing Eq. (\ref{tBdG}) to a new form  
\be
i\partial_{\tilde t}
\left(
\begin{array}{c}
\tilde u^+\\
\tilde u^-
\end{array}
\right)~=~
\left[
-2\frac{\tilde t}{\tilde\tau_Q} \sigma^x +
2i\sigma^v \partial_{\tilde n}+
\frac{4}{iv} \partial_{\tilde n}
\right]
\left(
\begin{array}{c}
\tilde u^+\\
\tilde u^-
\end{array}
\right)~.
\label{calH}
\ee
Here $\sigma^v=\sigma^y\sqrt{1-\frac{4}{v^2}}+\frac{2}{v}~\sigma^z$. Up to an unimportant 
rotation of a Pauli matrix $\sigma^y\to\sigma^v$ and the momentum-dependent energy shift 
$\frac{4}{iv}\partial_{\tilde n}$, the new Eq. (\ref{calH}) is a homogeneous version of 
the old Eq. (\ref{tBdG}), but with a longer effective quench time 
$\tilde\tau_Q ~=~\tau_Q~\left(1-\frac{4}{v^2}\right)^{-3/2}~>~\tau_Q$. 

Consequently, a quasimomentum representation
$( \tilde u^+ , \tilde u^- )~=~(a_k,b_k)~\exp(ik\tilde n-4ik\tilde t/v)/\sqrt{2\pi}$ 
brings the homogeneous Eq. (\ref{calH}) to the Landau-Zener form:
\be
i \frac{d}{ds}
\left(
\begin{array}{c}
a_k\\
b_k
\end{array}
\right)~=~
\frac12
\left[
~-\delta_k ~ s ~\sigma^x~+~\sigma^v~
\right]
\left(
\begin{array}{c}
a_k\\
b_k
\end{array}
\right)~,
\ee
where $s=k~\tilde t$ is a new time variable and $\delta_k=1/4k^2\tilde\tau_Q$
is a new transition rate. The Landau-Zener formula $p_k=\exp(-\pi/2\delta_k)$ gives 
excitation probability for a quasiparticle $k$ and density of excited quasiparticles is 
\be
d(v>2)~=~
\int_{-\Lambda}^\Lambda \frac{dk}{2\pi}~p_k~=~
\frac{\left(1-\frac{4}{v^2}\right)^{3/4}}{2\pi\sqrt{2\tau_Q}}~,
\label{dv}
\ee
where $\Lambda\simeq1$ is an ultraviolet cut-off.
The integral is accurate for $\tilde\tau_Q\gg1$. When $v\gg 2$ the density 
\be
d(v\gg 2)~\approx~\frac{1}{2\pi\sqrt{2\tau_Q}}~\equiv~d_{\rm KZM}
\ee 
is the same as the density after a homogeneous quench with the same $\tau_Q$, 
see Ref. \cite{Dziarmaga2005}, but when $v\to2^+$ then $d$ is suppressed below 
the ``homogeneous'' density $d_{\rm KZM}$ by the factor $(1-4/v^2)^{3/4}$.

%%%%%%%%%%%%%%%%%%%%%%%%%%%%%%%%%%%%%%%%%%%%%%%%%%%%%%%%%%%%%%%%%%%%%%%%%%%%%%%%%%%
\begin{figure}
\includegraphics[width=0.99\columnwidth,clip=true]{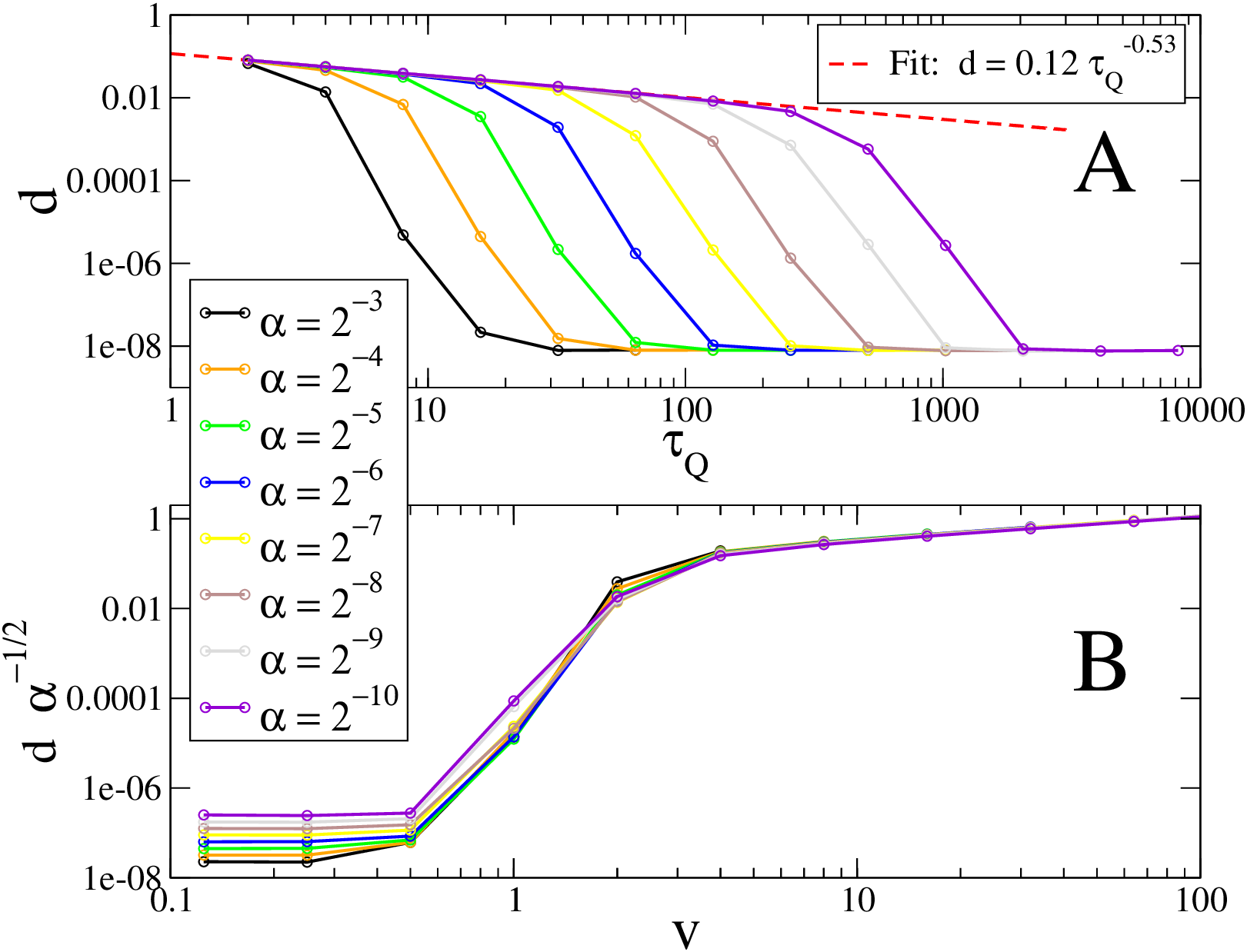}
\caption{ 
Numerical simulations of $N=400$ spins. In A final density of kinks $d(\tau_Q)$ 
for different slopes $\alpha$, and in B a rescaled final density $\alpha^{-1/2}~d(v)$. 
The solid lines are guide to the eye. The fit in panel A shows that when $v\gg2$ then 
$d\simeq\tau_Q^{-1/2}$ like in the homogeneous KZM, and when $v\ll2$ then $d$ is 
suppressed below the homogeneous KZM density.
}
\label{Fignumerics}
\end{figure}
%%%%%%%%%%%%%%%%%%%%%%%%%%%%%%%%%%%%%%%%%%%%%%%%%%%%%%%%%%%%%%%%%%%%%%%%%%%%%%%%%%%%%

%%%%%%%%%%%%%%%%%%%%%%%%%%%%%%%%%%%%%%%%%%%%%%%%%%%%%%%%%%%%%%%%%%%%%%%%%%%%%%%
\subsection{ Numerical results }
%%%%%%%%%%%%%%%%%%%%%%%%%%%%%%%%%%%%%%%%%%%%%%%%%%%%%%%%%%%%%%%%%%%%%%%%%%%%%%%

\begin{figure}
\includegraphics[width=0.99\columnwidth,clip=true]{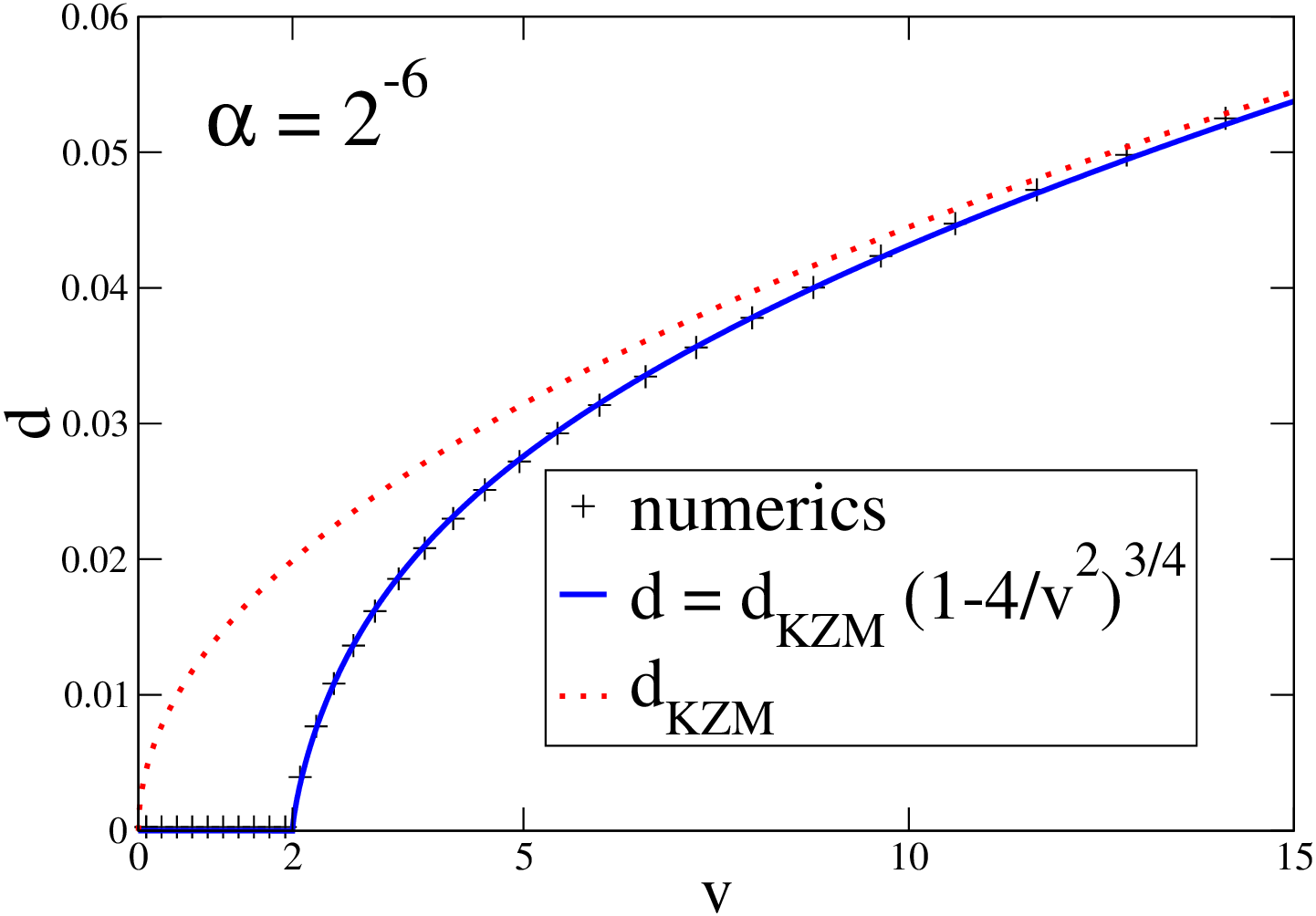}
\caption{ 
Comparison between Eq. (\ref{dv}) (solid blue), the homogeneous KZM (dotted red), 
and numerical simulations on a lattice of $N=1000$ spins (crosses) at a fixed 
slope $\alpha=2^{-6}$. 
}
\label{Figv}
\end{figure}

Since the long-wavelength Eq. (\ref{LWA}) does not give self-consistent 
long-wavelength solutions when $v\to2$, we simulated the
exact time-dependent version of the Bogoliubov-de Gennes equations 
\be 
i\frac{du_{n,m}^\pm}{dt}~=~2g_n(t)u_{n,m}^\mp~-~2u_{n\mp 1,m}^\mp~,
\label{exacttBdG} 
\ee
on a finite lattice of $N$ sites for a time-dependent transverse magnetic field
\be
g_n(t)~=~1~+~\tanh\left[\alpha(n-vt)\right]~
\label{slantv}
\ee
with a moving critical point at $n_c=vt$, compare Eq. (\ref{slant}) and 
Fig. \ref{Figslant}. Results are shown in Figs. \ref{Fignumerics} and \ref{Figv}. 
Figure \ref{Figv} demonstrates good quantitative agreement between Eq. (\ref{dv}) 
and numerical results, despite the breakdown of the long-wavelength approximation
near $v=2$.

%%%%%%%%%%%%%%%%%%%%%%%%%%%%%%%%%%%%%%%%%%%%%%%%%%%%%%%%%%%%%%%%%%%%%%%%%%%%%%%%%%%
%%%%%%%%%%%%%%%%%%%%%%%%%%%%%%%%%%%%%%%%%%%%%%%%%%%%%%%%%%%%%%%%%%%%%%%%%%%%%%%%%%%%%

%%%%%%%%%%%%%%%%%%%%%%%%%%%%%%%%%%%%%%%%%%%%%%%%%%%%%%%%%%%%%%%%%%%%%%%%%%%%%%%%%%%%%
\section{ Conclusion }
%%%%%%%%%%%%%%%%%%%%%%%%%%%%%%%%%%%%%%%%%%%%%%%%%%%%%%%%%%%%%%%%%%%%%%%%%%%%%%%%%%%%%

We made the general estimate 
Eqs. (\ref{hatv},\ref{hatvtauQ},\ref{hatvalpha},\ref{hattauQ}) when 
an inhomogeneous quench cannot be considered homogeneous with respect to KZM. Then 
we solved the problem in detail in the particular case of the quantum Ising chain where 
$z=1$ and the threshold velocity $\hat v=2$ is equal to velocity of quasiparticles at 
the critical point. Excitation of kinks is dramatically suppressed when a critical front
propagates slower than $\hat v$ and the ferromagnetic phase is able to communicate its 
choice of ferromagnetic polarization to the paramagnetic phase ahead of the front. In 
contrast, when the front is much faster than $\hat v$ the communication across the front 
is not efficient enough and kinks are excited as in the homogeneous KZM. However, even 
above $\hat v$ density of excited kinks is suppressed below the ``homogeneous'' KZM 
density by a factor $\left(1-\frac{\hat v^2}{v^2}\right)^{3/4}$ which is significantly 
less than $1$ when $v$ is close to $\hat v^+$. 

Thus the general estimates (\ref{hatv},\ref{hatvtauQ},\ref{hatvalpha},\ref{hattauQ}) 
are confirmed by our solution of the quantum Ising chain, but we leave their interesting 
implications when $z\neq1$ or in more than one dimension for future exploration. 

The estimates and the solution suggest that ``inhomogeneous'' adiabatic quantum 
computers can be more adiabatic than their ``homogeneous'' counterparts.

%%%%%%%%%%%%%%%%%%%%%%%%%%%%%%%%%%%%%%%%%%%%%%%%%%%%%%%%%%%%%%%%%%%%%%%%%%%%%%%%%%%%%
\section*{ Acknowledgements } 
%%%%%%%%%%%%%%%%%%%%%%%%%%%%%%%%%%%%%%%%%%%%%%%%%%%%%%%%%%%%%%%%%%%%%%%%%%%%%%%%%%%%%

We thank Wojciech Zurek for discussions. Work of J.D. and M.M.R. was supported by 
Polish Government research projects N202 175935 and N202 174335 respectively, and 
the Marie Curie ATK project COCOS (contract MTKD-CT-2004-517186).

%%%%%%%%%%%%%%%%%%%%%%%%%%%%%%%%%%%%%%%%%%%%%%%%%%%%%%%%%%%%%%%%%%%%%%%%%%%%%%%%%%%%%

\end{document}